\documentstyle[seceq,epsfig]{ptptex}

\def\siml{\lower4pt \hbox{$\buildrel < \over \sim$}}
\def\simg{\lower4pt \hbox{$\buildrel > \over \sim$}}
\def\Mesz{M\'esz\'aros~}
\def\Pacz{Paczy\'nski~}

\def\nsns{NS-NS~}
\def\bhns{BH-NS~}

\def\eps{\epsilon}
\def\varep{\varepsilon}

\def\Fnu{F_{\nu}}

\def\Omj{\Omega_j}
\def\Mbh{M_{bh}}
\def\msun{M_\odot}
\def\etal{{\it et~al.}}
\def\bec{\begin{center}}
\def\enc{\end{center}}
\def\beq{\begin{equation}}
\def\enq{\end{equation}}
\def\bea{\begin{eqnarray}}
\def\ena{\end{eqnarray}}

\newcommand{\boxsize}{0.89\textwidth}

\title{ Origin of Gamma Ray Bursters\footnote{Based on a review to appear in 
{\it Prog.Theor.Phys} S.136 (1999), {``Black Holes and Gravitational Waves - 
New Eyes in the 21st Century"}, Yukawa International Seminar'99, eds. 
T. Nakamura et al} }

\author{ P. {\sc \Mesz }} 

\inst{Dpt. of Astronomy \& Astrophysics,
Pennsylvania State University, \\ University Park, PA 16802, USA}

\recdate{
}

\abst{
Major advances have been made in the field of gamma-ray bursts in the last
two years. The successful discovery of X-ray, optical and radio afterglows
has made possible the identification of host galaxies at cosmological distances. 
The energy release inferred in these outbursts place them among the most energetic 
and violent events in the Universe. They are thought to be the outcome of a 
cataclysmic stellar collapse or compact stellar merger, leading to a 
relativistically expanding fireball, in which particles are accelerated at 
shocks and produce nonthermal radiation. The substantial agreement between 
observations and the theoretical predictions of the standard fireball shock 
model provide confirmation of the basic aspects of this scenario. New issues
being raised by the most recent observations  concern the amount and the nature
of the collimation of the outflow and its implications for the energetics, 
the production of prompt bright flashes at wavelenghts much longer than gamma-rays, 
the time structure of the afterglow, its dependence on the central engine or 
progenitor system behavior, and the role of the environment on the evolution of 
the afterglow.
}

\begin{document}

\maketitle

\section{Introduction}

Gamma-ray bursts (GRB) are brief pulses of gamma-rays which pierce, for a
brief period of tens of seconds, an otherwise pitch-black gamma-ray sky.
They are detected about once a day, and while they are on, they outshine
everything else in the gamma-ray sky, including the Sun. In fact, they are
the brightest explosions in the Universe. Until a few years ago, they left no
trace at any other wavelengths besides gamma-rays,
until in early 1997 the Italian-Dutch satellite Beppo-SAX suceeded in
providing accurate X-ray measurements that allowed, after a delay of some
hours to process the position. This made possible their follow-up with large
ground-based optical and radio telescopes, which paved the way for the
measurement of redshift distances, the identification of candidate host
galaxies, and the confirmation that they were at cosmological distances,
comparable to that of the most distant galaxies and quasars ever measured.
Since even at those tremendous distances they appear so bright, their energy
output needs to be stupendous. It is comparable to burning up the entire
mass-energy of the Sun in a few tens of seconds, or to emit over that same
period of time as much energy as our Milky Way does in a hundred years.
The current interpretation of this radiation is that this large amount
of energy. released in a very short time in a very small region, expands
in highly relativistic outflow, which undergoes both internal shocks producing 
gamma-rays, and later develops a blast wave and reverse shock, as it is
decelerated by interaction with the external medium. The remarkable thing
about this model is that it successfully predicted 
the expected observational properties of the afterglows which such bursts
leave at X-ray, optical and radio wavelengths, which can be observed over months.
This fireball shock model and the blast wave model of the ensuing afterglow
have become the leading paradigms for the current understanding of GRB.

GRBs were first discovered in 1974 by the Vela military satellites,
monitoring for nuclear explosions in violation of the Nuclear Test Ban Treaty.
When these mysterious gamma-ray flashes were first detected, which did not
come from the Earth's direction, the first suspicion (quickly abandoned)
was that they might be the product of an advanced extraterrestrial civilization.
Soon, however, it was realized that this was a new and extremely puzzling
cosmic phenomenon. For the next 25 years, only these brief gamma-ray flashes
were observed, which vanished too
soon to get an accurate angular position and hence left no trace, or so
it seemed. Gamma-rays are notoriously hard to focus, so no gamma-ray ``images"
exist to this day: they are just pin-points of gamma-ray light.
This led to a huge interest and to numerous conferences and publications on the
subject, as well as to a proliferation of theories. In one famous  review
article at the 1975 Texas Symposium on Relativistic Astrophysics,
no fewer than 100 different possible theoretical models of GRB were listed,
most of which could not be ruled out by the observations then available!


A major advance occurred in 1992 with the launch of the Compton Gamma-Ray
Observatory, whose superb results are summarized in a review by
Fishman \& Meegan \cite{fm95}. The all-sky survey from the BATSE
instrument showed that bursts were isotropically distributed, strongly
suggesting a cosmological, or possibly an extended galactic halo distribution, 
with essentially zero dipole and quadrupole components. The spectra are non-thermal, 
typically fitted in the MeV range by broken power-laws whose energy per decade 
$\nu\Fnu$ peak is in the range 50-500 KeV \cite{band93}, the power law sometimes 
extending to GeV energies \cite{hur94}. GRB appeared to leave no detectable
traces at other wavelengths, except in some cases briefly in X-rays.
%
\begin{figure}[ht]
\vspace*{-0.5cm}
\begin{center}
\begin{minipage}[t]{0.5\textwidth}
\epsfxsize=\boxsize
\epsfbox{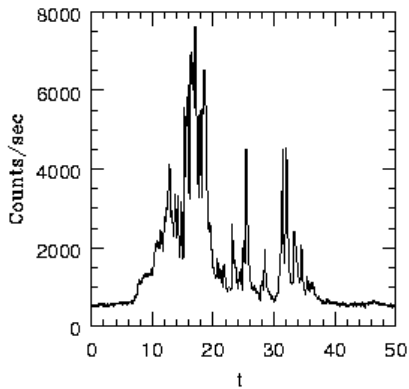}
\end{minipage}
\hspace{10mm}
\begin{minipage}[t]{0.4\textwidth}
\vspace*{-5cm}
\caption{\label{fig:batselc}
{\sc Time Profile of a Typical Gamma Ray Burst.}
The y-axis is the photon count rate in the 0.05-0.5 MeV energy,
the x-axis is the time in seconds after the trigger. Both before and
after the burst, no gamma-rays are detectable from the direction
of the burst (Fishman \& Meegan, 1995).
}
\end{minipage}
\end{center}
\vspace*{-0.5cm}
\end{figure}
The gamma-ray durations range from $10^{-3}$ s to about $10^3$ s, with a roughly
bimodal distribution of long bursts of $t_b \simg 2$ s and short bursts of
$t_b \siml 2$s \cite{kou93}, and substructure sometimes down to milliseconds.
The gamma-ray light curves range from smooth, fast-rise and quasi-exponential decay
(FREDs), through curves with several peaks, to highly variable curves with  many
peaks \cite{fm95,kou98}. The pulse distribution is complex \cite{pen96,nor98}, and
the time histories can provide clues for the geometry of the emitting regions
\cite{fen96,fen98}.
 
At cosmological distances the observed GRB fluxes imply enormous energies, 
and from causality they must arise in a small volume in a very short time, 
so an $e^\pm -\gamma$ fireball must form \cite{pac86,goo86,sp90}, which would expand
relativistically. The main difficulty with this was that a smoothly  expanding
fireball would  convert most of its energy into kinetic energy of accelerated
baryons (rather than photons), and would produce a quasi-thermal spectrum, while
the typical timescales would not explain events much longer than milliseconds.
This problem was solved with the introduction of the ``fireball shock model"
\cite{rm92,mr93a}, based on the realization that shocks are likely to arise,
at the latest when the fireball runs into an external medium, which would
occur after the fireball is optically thin and would reconvert the kinetic energy
into nonthermal radiation. The complicated light curves can be understood in terms
of internal shocks \cite{rm94} in the outflow itself, before it runs into the
external medium, caused by velocity variations in the outflow from the source,

The next major advance came in early 1997 when the Italian-Dutch satellite
Beppo-SAX succeeded in providing accurate X-ray measurements which, after a delay of
4-6 hours for processing, led to positions \cite{cos97},
allowing follow-ups at optical and other wavelengths, e.g. \cite{jvp97}. This paved
the way for the measurement of redshift distances, the identification of candidate
host galaxies, and the confirmation that they were indeed at cosmological distances
\cite{metz97,djo98_0703,kul98}. The detection of other GRB afterglows followed in
rapid succession, sometimes extending to radio \cite{fra97,fra98} and over timescales
of many months \cite{jvp98}, and in a number of cases resulted in the identification
of candidate host galaxies, e.g. \cite{sah97,bloo98_0508,ode98_1214}, etc.  The study
of afterglows has provided strong confirmation for the generic fireball shock model 
of GRB.  This model in fact led to a correct prediction \cite{mr97a}, in advance of 
the observations, of the quantitative nature of afterglows at wavelengths longer 
than $\gamma$-rays, which were in substantial agreement with the data
\cite{vie97a,tav97,wax97a,rei97,wrm97}.
 
A major issue raised by the large redshifts, e.g. \cite{kul98,kul99},
is that the measured $\gamma$-ray fluences imply a total energy of order
$10^{54}(\Omega_\gamma /4\pi)$ ergs, where $\Delta \Omega_\gamma$ is the solid
angle into which the gamma-rays are beamed. A beamed jet would clearly alleviate
the energy requirements, but it is only recently that tentative evidence has been
reported for evidence of a jet \cite{kul99,fru99,cas99}. Whether a jet is present or
not, such energies are possible \cite{mr97b} in the context of compact mergers
involving neutron star-neutron star (\nsns) or black hole-neutron star
(\bhns) binaries, or in hypernova/collapsar models involving a massive stellar
progenitor \cite{pac98,pop99}. In both cases, one is led to rely on MHD
extraction of the spin energy of a disrupted torus and/or a central
fast spinning BH, which can power a relativistic fireball resulting in the
observed radiation. 
 
While it is at present unclear whether there is a single or several classes
of GRB progenitors, there is a general consensus that they would all lead to the 
generic fireball shock scenario mentioned above. Two recent trends in the 
observations are the increasing hints pointing at a massive progenitor,
at least for those bursts where an afterglow is detected \cite{pac99},
and the possibility that, at least in some bursts, a peculiar supernova may
may be involved, e.g. \cite{whee99}.  Much of the current effort is dedicated to 
understanding the different possible progenitors more specifically, and trying 
to determine how this affects the details of the observable burst and afterglow.
Some of these progenitors, clearly, could be amenable also to observation with
the current generation of gravitational wave detectors, and the numerical
simulations of the collapse or merger share many interesting aspects with 
calculations motivated by gravitational wave searches.

\section{Progenitors and Black Hole Systems}
\label{sec:progen}

In the last few years it has become apparent that {\it most} plausible GRB 
progenitors would be expected to lead to a central black hole (BH) and a temporary 
debris torus around it. This includes both the currently popular massive
progenitor systems, such as hypernova or collapsars (including failed supernova Ib 
[SNe Ib], single or binary Wolf-Rayet [WR] collapse, etc. \cite{woo93,pac98})
\cite{pac91,woo93,mr97b,pac98,pop99},
NS-NS or NS-BH mergers \cite{pac91,mr97b}, Helium core - black hole
[He/BH] or white dwarf - black hole [WD-BH] mergers \cite{pop99,woo99}, 
accretion-induced collapse \cite{vs98,pfb99}, etc.. An important point is that the
overall energetics from these various progenitors do not differ by more than about
one order of magnitude \cite{mrw98b}. Another possibility is massive black holes
($\sim 10^3 - 10^5 \msun$) in the halos of galaxies.
Some related models involve a compact binary
or a temporarily rotationally stabilized neutron star, perhaps with a superstrong
field, e.g. \cite{us94,tho94,vs98,spr99}, which ultimately also should lead to a
BH plus debris torus.
 
The two large reservoirs of energy available in these systems are the binding
energy of the orbiting debris, and the spin energy of the black hole
\cite{mr97b}. The first can provide up to 42\% of the rest mass energy of
the disk, for a maximally rotating black hole, while the second can provide
up to 29\% of the rest mass of the black hole itself. However, as in the
related AGN case, the question is how to to extract this energy and how to deposit 
it at larger distances, where it produces optically thin radiation.
 
A classical energy extraction mechanisms is the $\nu\bar\nu \to e^+ e^-$ 
process \cite{eic89}, which can tap the thermal energy of the torus produced
by viscous dissipation. To be efficient, the neutrinos must escape before
being advected into the hole;
on the other hand, the efficiency of conversion into pairs (which scales with
the square of the neutrino density) is low if the neutrino production is too
gradual. Typical estimates suggest a  fireball of $\siml 10^{51}$ erg
\cite{ruf97,fw98,mcfw99}, except perhaps in the collapsar case where
\cite{pop99} estimate $10^{52.3}$ ergs for optimum parameters. If the fireball
is collimated into a solid angle $\Omj$ then of course the apparent
``isotropized" energy would be larger by a factor $(4\pi/\Omj)$ , but unless
$\Omj$ is $\siml 10^{-3}-10^{-4}$ this would fail to satisfy the apparent
isotropized energy of $4\times 10^{54}$ ergs deduced for GRB 990123
\cite{kul99}.

A more efficient mechanism for extracting and transporting the energy from
the vicinity of the BH may be through magnetic torques. Strong 
magnetic fields may be generated by the differential rotation in the torus 
\cite{pac91,napapi92,mr97b,ka97}, in any of the above progenitors.
In \nsns mergers, even before the BH forms the shrinking binary may lead to winding 
up of the fields and dissipation in the last stages before the merger 
\cite{mr92,vie97a}.  
An even larger energy source is provided by the spin of the hole itself, especially
if formed from a coalescing compact binary, which is likelier to be rotating 
close to the maximal rate. Since the BH is more massive, it could contain more 
energy than the torus. The energy extractable in principle through MHD coupling to 
the rotation of the hole by the B-Z (Blandford \& Znajek \cite{bz77}) effect could 
then be even larger than that contained in the orbiting debris \cite{mr97b,pac98}.
Collectively, any such MHD outflows have been referred to as Poynting jets.

The various progenitors differ only slightly in the mass of the BH; they can 
differ more on the debris torus mass, can differ even more markedly
in the amount of rotational energy contained in the BH. Strong magnetic
fields, of order $10^{15}$ G, are needed needed to carry away the rotational
or gravitational energy in a time scale of tens of seconds \cite{us94,tho94},
which may be generated on such timescales by a convective dynamo mechanism,
the conditions for which are satisfied in freshly collapsed neutron stars
or neutron star tori \cite{dt92,klurud98}.
If the magnetic fields do not thread the BH,
then a Poynting outflow can at most carry the gravitational binding energy
of the torus. For a maximally rotating BH this is 42\%, and for a slow-rotating 
BH this is 6\% of the torus rest mass, multiplied by an additional efficiency
factor $\eps$ for converting MHD energy into radiation.
The torus or disk mass in a NS-NS merger is\cite{ruja98} $M_d\sim
10^{-1}-10^{-2}\msun$ , and for a NS-BH, a He-BH, WD-BH merger or a binary
WR collapse it may be estimated at \cite{pac98,fw98} $M_d \sim 1\msun$.
In the HeWD-BH merger and WR collapse the mass of the disk is uncertain due
to lack of calculations on continued accretion from the envelope, so $1\msun$
is just a rough estimate. The maximum torus-based MHD energy extraction is
then
\beq
E_{max,t} \sim 1-10 \times 10^{53} \eps (M_d/\msun)~\hbox{ergs}~.
\label{eq:ediskcases}
\enq

If the magnetic fields in the torus thread the BH, the spin energy of the BH
can in principle be extracted via the \cite{bz77} (B-Z) mechanism
(\cite{mr97b}).  The extractable energy is
\beq
E_{max,bh}\sim
(f(a)/0.29) \eps \Mbh c^2 \sim 5\times 10^{53} f(a) \eps (\Mbh/\msun)~\hbox{ergs}.
\label{eq:ebh}
\enq
where $f(a)=1-([1+\sqrt{1-a^2}]/2 )^{1/2} \leq 0.29$ is the rotational efficiency 
factor, and $a = Jc/G M^2$ is the rotation parameter, which equals 1 for a 
maximally rotating black hole. The $f(a)$
rotational factor is is small unless $a$ is close to 1, where it rises sharply
to its maximum value $f(1)=0.29$, so the main requirement is a rapidly
rotating black hole, $a \simg 0.5$. For a massive progenitor it is then
imperative that the collapsed rotate fast, and hence a system with a compact
binary companion seems called for.
A binary fast-rotating WR scenario might do this, which does not differ much in 
its final details from the He-BH merger, depending on what fraction of the He 
core gets accreted along the rotation axis as opposed to along the 
equator \cite{fw98}.  For a fast rotating BH of $2.5-3\msun$ threaded by the 
magnetic field, the maximal energy carried out by the jet is then similar or 
somewhat larger than in the NS-NS case. Rapid rotation is also guaranteed in 
an NS-NS merger, since the radius is close to that of a black hole and the final
orbital spin period is close to the required maximal spin rotation period.
Some scenarios less likely to produce a fast rotating BH are the NS-BH merger
(where the rotation parameter could be limited to $a \leq M_{ns}/\Mbh$,
unless the BH is already fast-rotating), and the (single) failed SNe Ib, where the
last material to fall in would have maximum angular momentum, but the
material that was initially close to the hole has less angular momentum.
Some examples are discussed in the accompanying article by Ruffert \cite{ruf99}.
The main point that is worth stressing is that
the total energetics differ between the various models at most
by a factor 20 for a Poynting (MHD) jet powered by the torus binding energy,
and by a factor of a few for Poynting jets powered by the BH spin energy, 
depending on the rotation parameter. 
%

The major difference between the various models is expected to be in the
{\it location} where the burst occurs relative to the host galaxy
(see \S \ref{sec:env}). They are also likely to differ substantially in
the efficiency of producing a directly observable relativistic outflow, as well
as in the amount of collimation of the jet they produce. The conditions for the
efficient escape of a high-$\Gamma$ jet are less propitious if the ``engine" is
surrounded by an extensive envelope. In this case the jet has to ``punch
through" the envelope, and its ability to do so may be crucially dependent
on the level of viscosity achieved in the debris torus (e.g. \cite{mcfw99}),
higher viscosities leading to more powerful jets. 
The simulations, so far, are nonrelativistic and one can only infer that high enough 
viscosities can lead to jets capable of punching though a massive (several $\msun$)
envelope. This is facilitated, of course, if the envelope is fast-rotating,
as in this case there is a centrifugally induced column density minimum
along the spin axis, which might be small enough to allow punch-through
to occur. If they do, a very tightly collimated beam may arise.
``Cleaner" environments, such as NS-BH or NS-NS merger, or rotational
support loss/accretion induced collapse to BH would have much less
material to be pushed out of the way by a jet. In these, however, there is no 
natural choke to collimate a jet, which might therefore be somewhat wider than 
in a massive progenitor case.

\section{The Generic Fireball Shock Scenario}
\label{sec:fball}
 
No matter what the progenitor is, the ultimate result must almost unavoidably
be an $e^\pm,\gamma$ fireball, which initially will be very optically thick. 
The initial dimensions must be of order $c t_{var} \siml 10^7$ cm, since 
variability timescales are $t_{var}\simg 10^{-3}$ s.  Most of the spectral energy 
is observed above 0.5 MeV, hence the $\gamma\gamma \to e^\pm$ mean free path is 
very short. Due to the highly super-Eddington luminosity, this fireball must 
expand. Since many bursts show spectra extending above 1 GeV, the flow must be 
able to avoid degrading these via photon-photon interactions to energies
below the threshold $m_e c^2=$ 0.511 MeV\cite{hb94}. To avoid this, it seems 
inescapable that the flow must be expanding with a very high Lorentz factor 
$\Gamma$, since then the relative angle at which the photons collide is less 
than $\Gamma^{-1}$ and the threshold for the pair production is diminished. 
This condition is
\beq
\Gamma \simg 10^2 [(\eps_\gamma / {\rm 10 GeV}) (\eps_t /{\rm MeV}) ]^{1/2}~,
\enq
in order for photons $\eps_\gamma $ to escape annihilation
against target photons of energy $\eps_t \sim 1$ Mev \cite{m95,hb94}.
I.e., a {\it relativistically} expanding fireball is expected.
From general considerations \cite{mlr93}, an outflow arising
from an initial energy $E_o$ imparted to a mass $M_o << E_0/c^2$ within a
radius $r_l$ will lead to an expansion where initially the bulk Lorentz $\Gamma
\propto r$, while comoving temperature drops $\propto r^{-1}$. Since
$\Gamma$ cannot increase beyond $\Gamma_{max} \sim \eta \sim E_o/M_o c^2$,
which is achieved at a radius $r/r_l \sim \eta$, beyond this radius the flow
begins to coast, with $\Gamma \sim \eta \sim $ constant \cite{pac91}.
However, the observed $\gamma$-ray spectrum
observed is generally a broken power law, i.e., highly nonthermal. The
optically thick $e^\pm \gamma$ fireball cannot, by itself, produce such a
spectrum (it would tend rather to produce a modified blackbody,
\cite{pac86,goo86}). In addition,
the expansion would lead to a conversion of internal energy into kinetic
energy of expansion, so even after the fireball becomes optically thin,
it would be highly inefficient, most of the energy being in the kinetic
energy of the associated protons, rather than in photons.

The most likely way to achieve a nonthermal spectrum in an energetically
efficient manner is if the kinetic energy of the flow is re-converted into
random energy via shocks, after the flow has become optically thin \cite{rm92}.
Two different types of shocks may arise in this scenario. In the
first case (a) the expanding fireball runs into an external medium (the ISM, or a
pre-ejected stellar wind\cite{rm92,mr93a,ka94a,sapi95}. The second
possibility (b) is that \cite{rm94,px94}, even before external shocks occur,
internal shocks develop in the relativistic wind itself, faster portions of the
flow catching up with the slower portions.
This is a completely generic model, which is independent of the specific
nature of the progenitor, which has been successful in explaining the major 
observational properties of the gamma-ray emission, and is the main paradigm 
used for interpreting the GRB observations.

External shocks will occur in an impulsive outflow of total energy
$E_o$ in an external medium of average particle density $n_o$ at a radius
\beq
r_{dec} \sim (3E_o/4\pi n_o m_p c^2 \eta^2)^{1/3} \sim 
 10^{17} E_{53}^{1/3} n_o^{-1/3} \eta_2^{-2/3} ~{\rm cm}~,
\label{eq:rdec}
\enq
and on a timescale $t_{dec} \sim  r_{dec}/(c\Gamma^2)$ $\sim 3\times 10^2  
E_{53}^{1/3} n_o^{-1/3}\eta_2^{-8/3}$ s,
where $\eta=\Gamma = 10^2\eta_2$ is the final bulk Lorentz factor of the ejecta.
Variability on timescales shorter than $t_{dec}$
may occur on the cooling timescale or on the dynamic timescale for
inhomogeneities in the external medium, but generally this is not ideal for
reproducing highly variable profiles\cite{sapi98}.
However, it can reproduce bursts with several peaks\cite{pm98a}
and may therefore be applicable to the class of long, smooth bursts.
 
The same behavior $\Gamma \propto r$ with comoving temperature $\propto
r^{-1}$, followed by saturation $\Gamma_{max} \sim \eta$ at the same radius
$r/r_l \sim \eta$ occurs in a wind scenario \cite{pac90}, if one assumes
that a lab-frame luminosity $L_o$ and mass outflow $\dot M_o$ are injected
at $r\sim r_l$ and continuously maintained over a time
$t_w$; here $\eta=L_o/ {\dot M_o c^2}$. In such wind model, internal shocks
will occur at a radius \cite{rm94}
\beq
r_{dis} \sim  c t_{var} \eta^2 \sim 3\times 10^{14} t_{var} \eta_2^2 ~ {\rm cm},
\label{eq:rdis}
\enq
on a timescale $t_w \gg t_{var} \sim r_{dis}/(c\eta^2)$ s, 
where shells of different energies $\Delta \eta \sim \eta$ initially separated
by $c t_v $ (where $t_v \leq t_w$ is the timescale of typical variations in
the energy at $r_l$) catch up with each other.
In order for internal shocks to occur above
the wind photosphere $r_{ph} \sim {\dot M} \sigma_T /(4\pi m_p c \Gamma^2)$
$=1.2\times 10^{14} L_{53}\eta_2^{-3}$ cm, but also at radii greater than the
saturation radius (so that most of the energy does not come out in the
photospheric quasi-thermal radiation component) one needs to have
$7.5\times 10^1 L_{51}^{1/5} t_{var}^{-1/5} \siml \eta
3\times 10^2 L_{53}^{1/4} t_{var}^{-1/4}$.
This type of models have the advantage\cite{rm94} that they allow an
arbitrarily complicated light curve, the shortest variation timescale $t_{var}
\simg 10^{-3}$ s being limited only by the dynamic timescale at $r_l$, where
the energy input may be expected to vary chaotically.
Such internal shocks have been shown explicitly  to
reproduce (and be required by) some of the more complicated
light curves\cite{sapi98,kps98,pm99int} (see however \cite{dermit98}).

A potentially valuable diagnostic tool for the central engine of GRB
is the power density spectrum (PDS). An analysis of
a sample of 228 BATSE burst  light curves \cite{beloborodov+98}
comes up with two interesting results, namely the logarithmic slope of the PDS
between $10^{-2}$ and 2 Hz is approximately -5/3, and there is a cutoff of
the average PDS above 2 Hz. While these experimental results need further
verification by independent data analysis groups, the experience with radio and
accreting pulsars, black hole binaries and QPO sources indicates that time series
analyses can provide valuable constraints on physical source mechanisms, e.g. 
\cite{beloborodov99}.
\begin{figure}[ht]
\begin{center}
\begin{minipage}[t]{0.48\textwidth}
\epsfxsize=\boxsize
\epsfbox{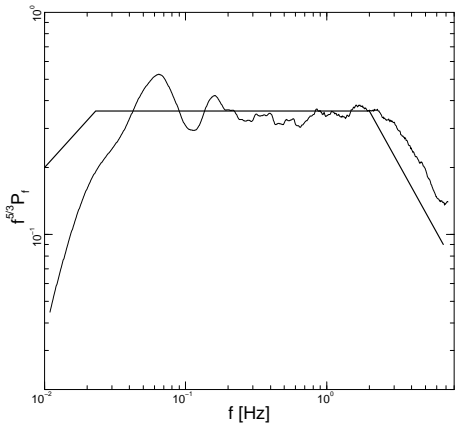}
\end{minipage}
\hspace{2mm}
\begin{minipage}[t]{0.4\textwidth}
\epsfxsize=\boxsize
\vspace*{-6cm}
\caption{\label{fig:pds}
Average power density spectrum $P_f$ of simulated bursts from internal shocks,
compared with the observed PDS (thick line) determined by \cite{beloborodov+98}.
The average is done over 112 bursts with a square-sine modulated wind of random 
period between $t_w/4$ and $t_w$, with a power-law luminosity distribution and a 
constant comoving rate density evolution satisfyng the observed logN-logP
constraints (Spada, Panaitescu \& \Mesz 1999).
}
\end{minipage}
\end{center}
\vspace*{-.5cm}
\end{figure}
Using a simple kinematical model for the shell ejection and
collision and assuming the BATSE radiation to be due to synchrotron and/or
IC scattered radiation, \cite{psm99,spm99} have calculated the
light curves and PDS expected for a range of total burst energies and for a
total mass ejected and bulk Lorentz factor distribution compatible with the
internal shock scenario (Figure \ref{fig:pds}). The distance distribution
also contributes to the observed PDS, and this is taken into account by
using a distribution compatible with the observed logN-logP counts.
For optically thin winds, a slope approaching -5/3 requires a non-random
Lorentz factor distribution, e.g. with an asymmetrical time modulation
so as to produce a larger number of collisions at low frequencies.
A cutoff at high frequencies ($\sim 2$ Hz) can be understood in
terms of shocks which increasingly occur below the scattering photosphere
of the outflow, or a deficit of energy in short pulses due to the modulation
of the Lorentz factors favoring shocks arising further out.

An interesting development is that significant fraction of bursts appear to have
low energy spectral slopes steeper than 1/3 in energy\cite{preece+98,crider+97}.
This has motivated consideration of a thermal or nonthermal\cite{liang+97,liang+99}
comptonization mechanism. While an astrophysical model where this mechanism 
would arise naturally has been  left largely unspecified, 
\begin{figure}[ht]
\begin{center}
\epsfbox{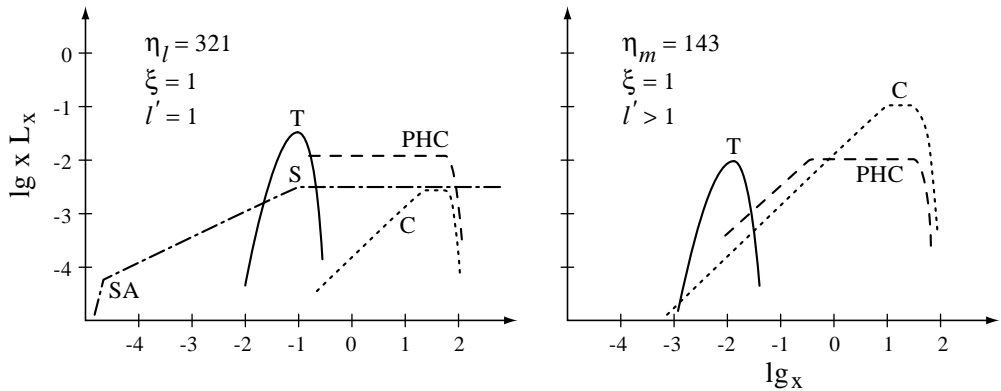}
\caption{\label{fig:photsp}
Luminosity per decade $xL_x$ vs. $x=h\nu/m_e c^2$ for two values of 
$\eta=L/{\dot M}c^2$ and marginal (left) or large (right) pair compactness.  
T: thermal photosphere, PHC: photospheric comptonized component; S: shock synchrotron; 
C: shock pair dominated comptonized component (\Mesz \& Rees, 1999b).}
\end{center}
\end{figure}
internal shocks which lead to pair formation\cite{ghiscel99apjl}
could have a self-regulated low pair temperature favoring comptonization.
There is also growing evidence that the apparent clustering of the break energy
of GRB spectra in the 50-500 keV range may not be due to observational selection
\cite{preece+98,brainerd+98peak,dermer+99apjl}. 
Models to explain this, e.g., through a Compton attenuation 
model\cite{brainerd+98apj}  
require reprocessing by an external medium whose column density adjusts itself to a 
few g cm$^{-2}$.  More recently a preferred break has been attributed to a blackbody
peak at the comoving pair recombination temperature in the fireball photosphere
\cite{eichlerlev99}.  
In this case a steep low energy spectral slope is
provided by the Rayleigh-Jeans part of the photosphere, and the high energy power
law spectra and GeV emission require a separate explanation. In order for such
photospheres to occur at the pair recombination temperature in the accelerating
regime requires an extremely low baryon load. For very large baryon loads, a
related explanation has been invoked \cite{tho94},
considerng scattering of photospheric photons off MHD turbulence in the coasting
portion of the outflow, which upscatters the adiabatically cooled photons
up to the observed break energy. These ideas have been synthesized \cite{mr99b}
to produce a generic scenario in which the presence
of a photospheric component as well as shocks subject to pair breakdown can produce
steep low energy spectra and preferred breaks.

\section{The Simple Standard Afterglow Model}
\label{sec:staaft}
 
The dynamics of GRB and their afterglows can be understood in a fairly simple
manner, independently of any uncertainties about the progenitor systems, using
a generalization of the method used to model supernova remnants. The simplest
hypothesis is that the afterglow is due to a relativistic expanding blast wave,
which decelerates as time goes on \cite{mr97a}.
The complex time structure of some bursts suggests that the central trigger may
continue for up to 100 seconds, the $\gamma$-rays possibly being due to
internal shocks. However, at much later times all memory of the initial time
structure would be lost: essentially all that matters is how much energy and
momentum has been injected; the injection can be regarded as instantaneous in
the context of the much longer afterglow.
As pointed in the original fireball shock paper \cite{rm92}, the external shock
bolometric luminosity builds up as $L\propto t^2$ and decays as $L\propto 
t^{-(1+q)}$.
At the deceleration radius (\ref{eq:rdec}) the fireball energy and the bulk Lorentz
factor decrease by a factor $\sim 2$ over a timescale $t_{dec}\sim
r_{dec}/(c\Gamma^2)$, and thereafter the bulk Lorentz factor decreases as a
power law in radius. This is
\beq
\Gamma \propto r^{-g}\propto t^{-g/(1+2g)}~,~r\propto t^{1/(1+2g)},
\label{eq:Gamma}
\enq
with $g$ is 3 for the radiative, 3/2 for the adiabatic regime (in which
$\rho r^3 \Gamma \sim$ constant or $\rho r^3 \Gamma^2 \sim$ constant).

The most obvious radiation mechanism is synchrotron, whose
peak frequency in the observer frame is $\nu_m \propto \Gamma B' \gamma^2$,
and both the comoving field $B'$ and electron Lorentz factor $\gamma$ are
expected to be proportional to $\Gamma$ \cite{mr93a}. As
$\Gamma$ decreases, so will $\nu_m$, and the radiation will move to longer
wavelengths. Looking at the implications of this for the forward blast wave,
\cite{pacro93,ka94b} discussed the possibility of detecting at late times a radio or
optical afterglow of the GRB. A more detailed treatment of the fireball dynamics
indicates that approximately equal amounts of energy are radiated by the forward
blast wave, moving with $\sim\Gamma$ into the surrounding medium, and by a
reverse shock propagating with $\Gamma_r -1 \sim 1$ back into the ejecta
\cite{mr93a}. The electrons are shocked to much higher energies in
the forward shock than in the reverse shock, producing a two-step synchrotron
spectrum which during the deceleration time $t_{dec}$ peaks in the optical
(reverse) and in the $\gamma/X$ (forward) \cite{mr93b,mrp94}. 
Detailed calculations and predictions of the time evolution of such a
forward and reverse shock afterglow model (\cite{mr97a}) preceded the
observations of the first afterglow GRB970228 (\cite{cos97,jvp97}), which
was detected in $\gamma$-rays, X-rays and several optical bands, and was
followed up for a number of months.
\begin{figure}[ht]
\begin{center}
\begin{minipage}[t]{0.5\textwidth}
\epsfxsize=\boxsize
\epsfbox{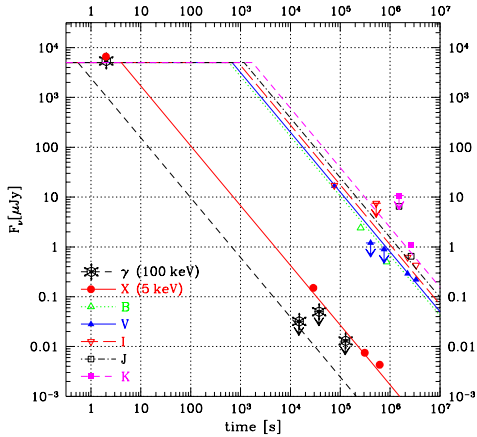}
\end{minipage}
\hspace{5mm}
\begin{minipage}[t]{0.3\textwidth}
\vspace*{-4cm}
\caption{\label{fig:0228_lightcurve}
The observed light curves of the afterglow of GRB 970228 at various wavelenghts,
compared \cite{wrm97} to the simple blast wave model predictions of \cite{mr97a}.
}
\end{minipage}
\end{center}
\end{figure}

The simplest spherical afterglow model concentrates on the
properties of the forward blast wave only. The flux at a given
frequency and the synchrotron peak frequency decay at a rate \cite{mr97a,mr99}
\beq
\Fnu\propto  t^{[3-2g(1-2\beta)]/(1+2g)}~~,~~\nu_m\propto t^{-4g/(1+2g)},
\label{eq:Fnu}
\enq
where $g$ is the exponent of $\Gamma$ (equ. [\ref{eq:Gamma}]) and $\beta$ is
the photon spectral energy slope. The decay rate of the forward shock $\Fnu$
in equ.(\ref{eq:Fnu}) is typically slower than that of the reverse shock
\cite{mr97a}, and the reason why the "simplest" model was stripped down to
its forward shock component only is that, for the first two years 1997-1998,
afterglows were followed in more detail only after the several hours needed by
Beppo-SAX to acquire accurate positions, by which time both reverse external
shock and internal shock components are expected to have become unobservable.
This simple standard model has been remarkably successful at explaining
the gross features and light curves of GRB 970228, GRB 970508 (after 2 days; for
early rise, see \S \ref{sec:postaft})
e.g. \cite{wrm97,tav97,wax97a,rei97} (see Figure \ref{fig:0228_lightcurve}).

\begin{figure}[ht]
\begin{center}
\vspace*{-1.cm}
\begin{minipage}[t]{0.4\textwidth}
\epsfxsize=\boxsize
\epsfbox{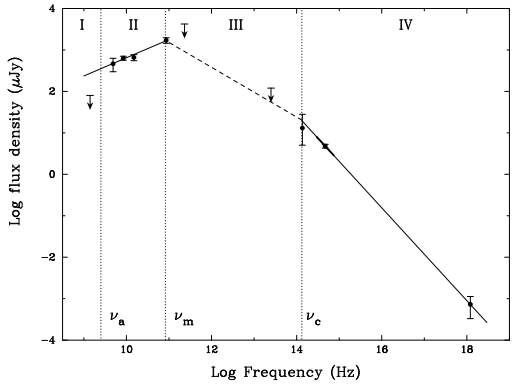}
\end{minipage}
\hspace{35mm}
\begin{minipage}[t]{0.3\textwidth}
\vspace*{-4cm}
\caption{\label{fig:0508_spec}
Snapshot spectrum of GRB 970508 at $t=12$ days and
 standard afterglow model fit \cite{wiga98}.
}
\end{minipage}
\vspace*{-1.cm}
\end{center}
\end{figure}

This standard afterglow model produces at any given time a three-segment power
law spectrum with two breaks. At low frequencies there is a steeply rising
synchrotron self-absorbed spectrum up to a self-absorption break $\nu_a$,
followed by a +1/3 energy index spectrum up to the synchrotron break $\nu_m$
corresponding to the minimum energy $\gamma_m$ of the power-law accelerated
electrons, and then a $-(p-1)/2$ energy spectrum above this break,
for electrons in the adiabatic regime (where $\gamma^{-p}$ is the electron
energy distribution above $\gamma_m$). A fourth segment and a third break is
expected at energies where the electron cooling time becomes short compared
to the expansion time, with a spectral slope $-p/2$ above that. With
this third ``cooling" break $\nu_b$, first calculated in \cite{mrw98} and
more explicitly detailed in \cite{spn98}, one has what has come to be called
the simple ``standard" model of GRB afterglows. One of the predictions of this
model \cite{mr97a} is that the relation between the temporal decay index $\alpha$,
for $g=3/2$ in $\Gamma\propto r^{-g}$, is related to the photon spectral energy
index $\beta$ through
\beq
\Fnu \propto t^\alpha \nu^\beta~~,\hbox{with}~~\alpha=(3/2)\beta~.
\label{eq:alphast}
\enq
This relationship appears to be valid in many (although not all) cases, especially
after the first few days, and is compatible with an electron spectral index $p\sim
2.2-2.5$ which is typical of shock acceleration, e.g. \cite{wax97a,spn98,wiga98},
etc.  As the remnant expands the photon spectrum moves to lower frequencies, and
the flux in a given band decays as a power law in time, whose index can change
as breaks move through it.
For the simple standard model, snapshot overall spectra have been deduced
by extrapolating spectra at different wavebands and times using assumed
simple time dependences \cite{wax97b,wiga98}. These can be used to derive rough
fits for the different physical parameters of the burst and
environment, e.g. the total energy $E$, the magnetic and electron-proton
coupling parameters ${\eps}_B$ and ${\eps}_e$ and the external density $n_o$
(see Figure \ref{fig:0508_spec}).

The simplest afterglow model is based on the following assumptions:
a) A single value of $E_o$ and $\Gamma_o=\eta$ is used,
b) the external medium $n_{ext}$ is homogeneous,
c) the accelerated electron spectral index $p$, the magnetic field and electron
to proton equipartion ratios $\varep_B$ and $\varep_e$ do not change in time,
d) the expansion is relativistic and the dynamics are given by $\Gamma \propto
r^{-3/2}$ (adiabatic), d) the outflow is spherical, e) the observed radiation 
is characterized by the scaling relations along the line of sight.
These assumptions, even if correct over some range, clearly would break down after
some time.  Estimates for the time needed to reach the non-relativistic expansion
regime are typically $\siml$ month(s) (\cite{vie97a}), or less if there is an initial
radiative regime $\Gamma\propto r^{-3}$.  However, even when electron radiative times
are shorter than the expansion time, it is unclear whether a regime $\Gamma\propto
r^{-3}$ should occur, since it would require strong electron-proton coupling
\cite{mrw98}. As far as sphericity, the standard model can be straightforwardly
generalized to the case where the energy is assumed to be channeled initially into a
solid angle $\Omj < 4\pi$ \cite{mlr93}. In this case \cite{rho97,rho99} a change
occurs after $\Gamma$ drops below $\Omj^{-1/2}$, after which the side of the jet
becomes observable, and soon thereafter one expects a faster decay of $\Gamma$
if the jet starts to expands sideways, leading to a decrease in the brightness.
A calculation based on the sideways expansion, using the usual scaling laws for a
single central line of sight \cite{rho99} leads then to a steepening of
the light curve. Until recently, no evidence for a steepening could be found
in afterglows over several months. E.g., in GRB 971214 \cite{ram98}, a
snapshot standard model fit and the lack of a break in the late light curve
could be, in principle, interpreted as evidence for lack of a jet, leading
to an (isotropic) energy estimate of $10^{53.5}$ ergs. While such large energy
outputs are possible in {\it either} NS-NS, NS-BH mergers \cite{mr97b} or in
hypernova/collapsar models \cite{pac98,pop99} using MHD extraction of the
spin energy of a disrupted torus and/or a central fast spinning BH, it is worth
stressing that what these snapshot fits constrain is only the {\it energy per
solid angle} \cite{mrw98b}. Also, the expectation of a break after some weeks
or months (e.g., due to $\Gamma$ dropping either below a few, or below
$\Omega_j^{-1/2}$) is based upon the simple impulsive (angle-independent delta
or top-hat function) energy input approximation. The latter is useful, but
departures from it would be natural, and certainly not surprising. In fact,
as discussed below, tentative evidence for beaming in one obejct has recently
been reported \cite{kul99,fru99,cas99}, but it is difficulty to ascertain, and
could be masked by a number of commonly expected effects.

\section{More Realistic (``Post-standard") Afterglow Models}
\label{sec:postaft}
 
Despite the success of the simple standard model, realistically one
could expect any of several fairly natural departures from this to occur. 
One of these is that the emitting region, as seen by the observer, should resemble 
a ring \cite{wax97b,pm98b,sari98}. This effect may,
in fact, be important in giving rise to the radio scintillation
pattern seen in several afterglows, since this requires the emitting source
to be of small dimensions, which is aided if the emission is ring-like,
e.g. in the example of GRB 970508 \cite{wkf98}. Another likely complication is
that the afterglows should show a diversity in their decay rates, not only due to
different $\beta$ but also from the possibility of a non-standard relation between
the temporal decay index $\alpha$ and the spectral energy index $\beta$, different
from equ. (\ref{eq:alphast}). 

The most obvious departure from the simplest
standard model occurs if the external medium is inhomogeneous: for instance, for
$n_{ext} \propto r^{-d}$, the energy conservation condition is $\Gamma^2 r^{3-d}
\sim$ constant, which changes significantly the temporal decay rates \cite{mrw98}.
Such a power law dependence is expected if the external medium is a wind, say from
an evolved progenitor star as implied in the hypernova scenario (such winds are
generally used to fit supernova remnant models). Another obvious non-standard effect,
which it is reasonable to expect, is departures from a simple impulsive injection
approximation (i.e. from a delta or top hat function with a single value for
$E_o$ and $\Gamma_o$). An example is if the mass and energy injected during the
burst duration $t_w$ (say tens of seconds) obeys $M(>\Gamma) \propto
\Gamma^{-s}$, $E(>\Gamma)\propto \Gamma^{1-s}$, i.e. more energy emitted with
lower Lorentz factors at later times (but still shorter than the gamma-ray pulse
duration). This would drastically change the temporal decay rate and extend the
afterglow lifetime in the relativistic regime, providing a late ``energy refreshment"
to the blast wave on time scales comparable to the afterglow time scale
\cite{rm98}. These two cases lead to a decay rate
\beq
\Gamma \propto r^{-g} \propto \cases{
  r^{-(3-d)/2} & ~; $n_{ext}\propto r^{-d}$;\cr
  r^{-3/(2+s)} & ~; $E(>\Gamma)\propto \Gamma^{1-s}$.\cr }
\label{eq:Gammanonst}
\enq
Expressions for the temporal decay index $\alpha (\beta,s,d)$ in $\Fnu\propto
t^\alpha$ are given by \cite{mrw98,rm98}, which now depend also on $s$ and/or $d$
(and not just on $\beta$ as in the simple standard relation of equ.(\ref{eq:alphast}).
The result is that the decay can be flatter (or steeper, depending on $s$ and $d$)
than the simple standard $\alpha= (3/2)\beta$.
A third non-standard effect, which is entirely natural, occurs when the energy
and/or the bulk Lorentz factor injected are some function of the angle. A simple case
is $E_o\propto \theta^{-j}$, $\Gamma_o\propto \theta^{-k}$ within a range of angles;
this leads to the outflow at different angles shocking at different radii and its
radiation arriving at the observed at different delayed times, and it has a marked
effect on the time dependence of the afterglow \cite{mrw98}, with $\alpha=\alpha
(\beta,j,k)$ flatter or steeper than the standard value, depending on $j,k$.
Thus in general, a temporal decay index which is a function of more than one
parameter
\beq
\Fnu\propto t^\alpha\nu^\beta~~,\hbox{with}~~\alpha=\alpha (\beta,d,s,j,k,\cdots )~,
\label{eq:alphanonst}
\enq
is not surprising; what is more remarkable is that, in many cases, the simple
standard relation (\ref{eq:alphast}) is sufficient to describe the gross overall
behavior at late times.
 
Strong evidence for departures from the simple standard model is provided by,
e.g., sharp rises or humps in the light curves followed by a renewed decay,
as in GRB 970508 (\cite{ped98,pir98a}). Detailed time-dependent model fits
\cite{pmr98} to the X-ray, optical and radio light curves of GRB 970228 and
GRB 970508 show that, in order to explain the humps, a {\it non-uniform} injection
or an {\it anisotropic} outflow is required.
%
These fits indicate that
the shock physics may be a function of the shock strength (e.g. the electron
index $p$, injection fraction $\zeta$ and/or $\epsilon_b,~\epsilon_e$ change
in time), and also indicate that dust absorption is needed to simultaneously
fit the X-ray and optical fluxes. The effects of beaming (outflow within a
limited range of solid angles) can be significant \cite{pmjet99}, but are coupled
with other effects, and a careful analysis is needed to disentangle them.
 
One consequence of ``post-standard" decay laws (e.g. from density inhomogeneities,
non-uniform injection or anisotropic outflow) is that the transition to a
steeper jet regime $\Gamma < \theta_j^{-1} \sim$ few can occur as late as six
months to a year after the outburst, depending on details of the energy input.
This transition is made more difficult to detect by the fact that, as numerical
integration over angles of the ring-like emission \cite{pmring98} show, the
transition is very gradual and the effects of sideways expansion effects are not
so drastic as inferred \cite{rho99} from the scaling laws along the central line of
sight. This is because even though the flux from the head-on part of the remnant
decreases faster, this is more than compensated by the increased emission
measure from sweeping up external matter over a larger angle, and by the
fact that the extra radiation, arising at larger angles, arrives later and
re-fills the steeper light curve. The inference (e.g. \cite{ram98,rho99}) that
GRB 970508 and a few other bursts were isotropic due to the lack of an observable
break is predicated entirely on the validity of the {\it simplest standard} fireball
assumption. Since these assumptions are drastic simplifications, and physically
plausible generalizations lead to different conclusions, one can interpret the
results of \cite{ram98,rho99} as arguments indicating that {\it post-standard}
features are, in fact, necessary in some objects.

\section{Prompt multi-wavelength flashes, reverse shocks and jets}
\label{sec:promptjet}
 
Prompt optical, X-ray and GeV flashes from reverse and forward shocks, as well as
from internal shocks, have been calculated in theoretical fireball shock models
for a number of years \cite{mr93b,mrp94,pm96,mr97a,sp99a}, as have been jets (e.g.
\cite{mr92,mlr93,mrp94}, and in more detail \cite{rho97,pmr98,pmjet99,rho99}).
Thus, while in recent years they were not explicitly part of the ``simple standard"
model, they are not strictly ``post"-standard either, since they generally use
the ``standard" assumptions, and they have a long history.
However, observational evidence for these effects were largely lacking, until
the detection of a prompt (within 22 s) optical flash from GRB 990123 with
ROTSE by \cite{ake99}, together with X-ray, optical and radio follow-ups
cite{kul99,gal99,fru99,and99,cas99,hjo99}. GRB 990123 is so far unique not only
for its prompt optical detection, but also by the fact that if it were emitting
isotropically, based on its redshift $z=1.6$ \cite{kul99,and99} its energy would
be the largest of any GRB so far, $4\times 10^{54}$ ergs. It is, however, also
the first (tentative) case in which there is evidence for jet-like emission
\cite{kul99,fru99,cas99}. An additional, uncommon feature is that a radio afterglow
appeared after only one day, only to disappear the next \cite{gal99,kul99}.
 
The prompt optical light curve of GRB 990123 decays initially as $\propto t^{-2.5}$
to $\propto t^{-1.6}$ \cite{ake99}, much steeper than the typical $\propto
t^{-1.1}$ of previous optical afterglows detected after several hours.
However, after about 10 minutes its decay rate moderates, and appears to
join smoothly onto a slower decay rate $\propto t^{-1.1}$ measured with
large telescopes \cite{gal99,kul99,fru99,cas99} after hours and days. The prompt
optical flash peaked at 9-th magnitude after 55 s \cite{ake99}, and in fact a
9-th magnitude prompt flash with a steeper decay rate had been predicted more than
two years ago \cite{mr97a}, from the synchrotron radiation of the reverse shock
in GRB afterglows at cosmological redshifts (see also {\cite{sp99a}). An optical
flash contemporaneous with the $\gamma$-ray burst, coming from the reverse shock
and with fluence corresponding to that magnitude, had also been predicted earlier
\cite{mr93b,mrp94}. An origin of the optical prompt flash in internal shocks
\cite{mr97a,mr99} cannot be ruled out yet, but is less likely since the optical
light curve and the $\gamma$-rays appear not to correlate well \cite{sp99b,gal99}
(but the early optical light curve has only three points). The subsequent slower
decay agrees with the predictions for the forward component of the external shock
\cite{mr97a,sp99b,mr99}.
 
The evidence for a jet is possibly the most exciting, although must still
be considered tentative. It is based on an apparent steepening of the light curve
after about three days \cite{kul99,fru99,cas99}. This
is harder to establish than the decay of the two previous earlier portions of
the light curve, since by this time the flux has decreased to a level where
the detector noise and the light of the host galaxy become important. However,
after correcting for this, the r-band data appears to steepen significantly.
(In the K-band, where the noise level is higher, a steepening is not obvious,
but the issue should be settled with further Space Telescope observations).
If real, this steepening is probably due to the transition between early
relativistic expansion, when the light-cone is narrower than the jet opening,
and the late expansion, when the light-cone has become wider than the jet,
leading to a drop in the effective flux \cite{rho97,kul99,mr99,rho99}. A rough
estimate leads to a jet opening angle of 3-5 degrees, which would reduce the total
energy requirements to about $4\times 10^{52}$ ergs. This is about two order of
magnitude less than the binding energy of a few solar rest masses, which, even
allowing for substantial inefficiencies, is compatible with currently favored
scenarios (e.g. \cite{pop99,mcfw99}) based on a stellar collapse or a compact
binary merger.
 
\section{Location and Environmental Effects}
\label{sec:env}
 
The location of the afterglow relative to the host galaxy center can
provide clues both for the nature of the progenitor and for the external
density encountered by the fireball. A hypernova model would be expected
to occur inside a galaxy in a high density environment $n_o > 10^3-10^5$ cm$^{-3}$.
Most of the detected and well identified afterglows are inside the projected image
of the host galaxy \cite{bloo98rome}, and some also show evidence for a dense
medium at least in front of the afterglow (\cite{ow98}).
For a number of bursts there are constraints from the
lack of a detectable, even faint, host galaxy \cite{sch98}, but at least for
Beppo-SAX bursts (which is sensitive only to long bursts $t_b \simg 20$ s) the
success rate in finding candidate hosts is high.

In NS-NS mergers one would expect a BH plus debris torus system and
roughly the same total energy as in a hypernova model, but the mean distance
traveled from birth is of order several Kpc \cite{bsp99}, leading to a burst
presumably in a less dense environment. The fits of \cite{wiga98} to the
observational data on GRB 970508 and GRB 971214 in fact suggest external densities
in the range of $n_o=$ 0.04--0.4 cm$^{-3}$, which would be more typical of a
tenuous interstellar medium. These could be within the volume of the galaxy,
but for NS-NS on average one would expect as many GRB inside as outside. This is
based on an estimate of the mean NS-NS merger time of $10^8$ years; other estimated
merger times (e.g. $10^7$ years, \cite{vdh92}) would give a burst much closer
to the birth site. BH-NS mergers would also occur in timescales $\siml 10^7$
years, and would be expected to give bursts well inside the host galaxy
(\cite{bsp99}; see however \cite{fw98}). In at least one ``snapshot" standard
afterglow spectral fit for GRB 980329 \cite{reila98} the deduced external
density is $n_o\sim 10^3$ cm$^{-3}$. In some of the other detected afterglows
there is other evidence for a relatively dense gaseous environments, as
suggested, e.g. by evidence for dust \cite{rei98} in GRB970508,
the absence of an optical afterglow and presence of strong soft X-ray
absorption \cite{gro97,mur97} in GRB 970828, the lack an an optical
afterglow in the (radio-detected) afterglow (\cite{tay97}) of GRB980329, and
spectral fits to the low energy portion of the X-ray afterglow of several
bursts \cite{ow98}. The latter observations may be suggestive of hypernova
models \cite{pac98,fw98}, involving the collapse of a massive star or its
merger with a compact companion.
 
One important caveat is that all afterglows found so far are based on Beppo-SAX
positions, which is sensitive only to long bursts $t_b \simg 20$ s \cite{hur98}.
This is significant, since it appears
likely that NS-NS mergers lead \cite{mcfw99} to short bursts with $t_b \siml 10$ s.
To make sure that a population of short GRB afterglows is not being
missed will probably need to await results from HETE \cite{hetepage} and from the
planned Swift \cite{swiftpage} mission, which is designed to accurately locate
300 GRB/yr.
 
An interesting case is the apparent coincidence of GRB 980425 with the
unusual SN Ib/Ic 1998bw \cite{gal98_SN}, which may represent a new class of SN
\cite{iwa98,bloomSN98}. If true, this could imply that some or perhaps
all GRB could be associated with SN Ib/Ic \cite{wawe98}, differring only in
their viewing angles relative to a very narrow jet. Alternatively,
the GRB could be (e.g. \cite{wes98}) a new subclass of GRB with
lower energy $E_\gamma \sim 10^{48} (\Omj /4\pi )$ erg, only rarely observable,
while the great majority of the observed GRB would have the energies $E_\gamma
\sim 10^{54}(\Omj/4\pi)$ ergs as inferred from high redshift observations.
The difficulties are that it would require extreme collimations
by factors $10^{-3}-10^{-4}$, and the statistical association is so far not
significant \cite{kip98}. However, two more GRB light curves may have been
affected by an anomalous SNR (see, e.g. the review of \cite{whee99}).

\begin{figure}[ht]
\begin{center}
\vspace*{-0.5cm}
\begin{minipage}[t]{0.6\textwidth}
\epsfxsize=\boxsize
\epsfbox{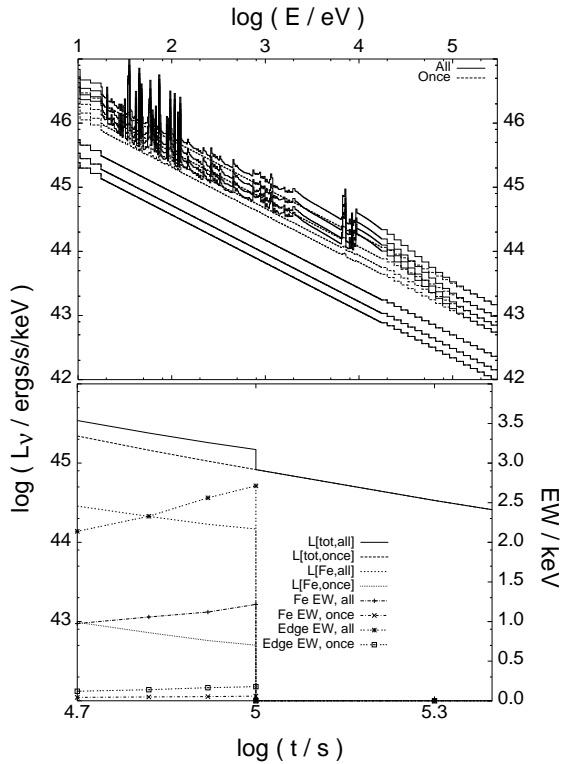}
\end{minipage}
\hspace{5mm}
\begin{minipage}[t]{0.29\textwidth}
\vspace*{-8cm}
\caption{\label{fig:shellsp}
Spectrum (top) of a hypernova funnel model for 
various observer times, showing (bottom) the total and Fe light curves and equivalent 
widths (Weth, \Mesz, Kallman \& Rees 1999), for with $R=1.5\times 10^{16}$ cm, 
$n=10^{10}$ cm$^{-3}$, and Fe abundance $10^2$ times solar.}
\end{minipage}
\vspace*{-0.5cm}
\end{center}
\end{figure}
The environment in which a GRB occurs should also influence the nature of the
afterglows in other ways. The blast wave and reverse shock that give rise to the
X-rays, optical, etc occur over timescales proportional to $t_{dec} \propto
n_{ext}^{-1/3}$ (equ.[\ref{eq:rdec}]) which is longer in lower density environments,
so for the same energy the flux is lower, roughly $\Fnu \propto E_o n_{ext}^{1/2}$,
contributing also to make afterglows in the intergalactic medium harder to detect.
However, in addition to affecting broad-band fluxes, one may also expect specific
spectral signatures from the external medium imprinted in the X-ray and optical
continuum, such as atomic edges and lines \cite{bkt97,pl98,mr98b}. These may be
used both to diagnose the chemical abundances and the ionization state (or local
separation from the burst), as well as serving as potential alternative redshift
indicators. (In addition, the outflowing ejecta itself may also contribute
blue-shifted edge and line features, especially if metal-rich blobs or filaments are
entrained in the flow from the disrupted progenitor debris \cite{mr98a}, which
could serve as diagnostic for the progenitor composition and outflow Lorentz factor).
To distinguish between progenitors (\S \ref{sec:progen}), an interesting prediction
(\cite{mr98b}; see also \cite{ghi98,bot98}) is that the presence of a measurable
Fe  K-$\alpha$ X-ray {\it emission} line could be a diagnostic of a hypernova,
since in this case one may expect a massive envelope at a radius comparable to a
light-day where $\tau_T \siml 1$, capable of reprocessing the X-ray continuum by
recombination and fluorescence. Detailed radiative transfer calculations have
been performed to simulate the time-dependent X/UV line spectra of massive
progenitor (hypernova) remnants\cite{weth+99}, see Figure \ref{fig:shellsp}.
Two groups \cite{piro98b,yosh98} have in fact
recently reported the possible detection of Fe emission lines in GRB 970508 and
GRB 970828.
 
\section{ Conclusions }
\label{sec:concl}
 
Gamma Ray bursts provide perhaps the most direct method for studying strong
gravity phenomena near black holes, and with over 4,000 bursts detected, they
would represent the most abundant supply of high quality photon data on black holes.
The fireball shock model of gamma-ray bursts has proved quite robust in
providing a consistent overall interpretation of the major features of these
objects at various frequencies and over timescales ranging from the short
initial burst to afterglows extending over many months. The standard internal
shock scenario is able to reproduce the properties of the $\gamma$-ray light
curves, while external shocks involving a forward blast wave and a reverse shock
are successful in reproducing the afterglows observed in X-rays, optical and radio.
The ``simple standard model" of afterglows, involving four spectral slopes and
three breaks is quite useful in understanding the `snapshot' multiwavelength
spectra of most afterglows. However, the effects associated with a jet-like
outflow and the possible differential beaming at various energies requires
further investigations, both theoretical and observational. 
Caution is required
in interpreting the observations on the basis of the simple standard model. For
instance, more detailed numerical models, as opposed to the more common analytical
scaling law models, show that the contributions of radiation from different angles
and the gradual transition between different dynamical and radiative regimes
lead to a considerable rounding-off of the spectral shoulders and light-curve
slope changes.
Time-dependent multiwavelength fits \cite{pmr98} of some
bursts also indicate that the parameters characterizing the shock physics change
with time, and a non-standard relation between the spectral and temporal decay 
slope observed in several objects, e.g. GRB 990123 \cite{kul99}, 
may provide evidence \cite{mr99} for ``post-standard" effects in such bursts.
 
Significant progress has been made in understanding how gamma-rays can arise in
fireballs produced by brief events depositing a large amount of energy in a
small volume, and in deriving the generic properties of the long wavelength
afterglows that follow from this.  There still remain a number of mysteries,
especially concerning the identity of their progenitors, the nature of the
triggering mechanism, the transport of the energy, the time scales involved,
and the nature and effects of beaming. However, even if we do not yet
understand the details of the gamma-ray burst central engine, it is clear that
these phenomena are among the most powerful transients in the Universe, and they
could serve as powerful beacons for probing the high redshift ($z > 5$)
universe. The modeling of the burst mechanism itself, as well as the resulting
outflows and radiation, will continue to be a formidable challenge to theorists
and to computational techniques. Nonetheless, the collective theoretical and 
observational understanding is vigorously advancing, and with dedicated new and 
planned observational missions under way, further significant progress  may be 
expected in the near future.

\section*{Acknowledgement}
I am grateful to Martin Rees for stimulating collaborations, to Alin
Panaitescu, Maddalena Spada and Chriss Weth for discussions, and Takashi Nakamura
and the Yukawa Institute for a stimulating meeting. Research supported by NASA
NAG-5 2857.



\end{document}